\documentclass[preprint,12pt]{elsarticle}
\usepackage[utf8]{inputenc}
\usepackage{graphicx}
\usepackage{amsmath}
\usepackage{enumerate}
\newcommand{\avg}[1]{\left \langle #1  \right\rangle}

\begin{document}

\title{Phase transition induced by traffic lights on a single lane road}
\author[famaf]{A. Chacoma}
\ead{achacoma@famaf.unc.edu.ar}
\author[fiestin,ib]{M. N. Kuperman}
\ead{kuperman@cab.cnea.gov.ar}
\author[fiestin,ib]{G. Abramson}
\ead{abramson@cab.cnea.gov.ar}

\address[famaf]{Instituto de F\'isica Enrique Gaviola (IFEG-CONICET), Ciudad Universitaria, 5000 C\'ordoba, Argentina}
\address[fiestin]{Centro At\'{o}mico Bariloche and CONICET, R8402AGP Bariloche, Argentina}
\address[ib]{Instituto Balseiro, Universidad Nacional de Cuyo, R8402AGP Bariloche, Argentina}

\date{}

\begin{abstract}
In this work we study the effect of a traffic light system on the flow of a single lane road by proposing  a traffic model based on a cellular 
automaton that also includes behavioral considerations.  We focus on the macroscopic characterization of the system  by studying the  changes in vehicle density and the occurrence of jams.  In this 
context we observe and characterize a phase transition between the free flow and jammed states. This transition is induced by the instabilities 
originated  by the vehicles stopping at the traffic lights. Moreover, we analyze the effect of these instabilities on the critical density of 
vehicles at which the transition occurs as a function of two parameters: (i) the in-flow of cars, (ii) the drivers behavior. For the latter we observe 
that the traffic light perturbations feedback on the drivers behavior can lead the system to different scenarios, which are also analyzed.
\end{abstract}

\maketitle

\section{Introduction}

More than 60\% of the world's population lives in urban  areas. Consequently, the logistics needed to mobilize all these  people around the cities in the context of their daily routines, has become a current topic of study 
\cite{bettencourt2010unified, WorldBank2012}. There are several problems related to urban 
transportation systems \cite{black2010sustainable}:
traffic jams, contamination and environmental problems,
generalized stress in the population, loss of time and in
consequence loss of money, etc.~\cite{small2013urban,mccarthy2012urbanization}.
Currently, most of the cities have an infrastructure aimed
at minimizing the adverse effects of traffic  and improving the overall well-being. 
However, the coordination of these resources becomes a complex problem as the size of
the cities increases and the system needs to be organized into several subsystems working independently,
isolated, without either connection or feedback among them.

Generally, in order to reduce
the problems related to logistic transportation, 
governmental strategies aim to increment the 
infrastructure of the system,
for instance building new roads, highways, bridges or
tunnels which are not always effective~\cite{murchland1970braess}.
On the other hand, the economic resources of the cities
are not always enough for this kind of infrastructure 
projects. Under such circumstances it is valid to ask if it would not be more efficient 
and less expensive to optimize the use of the
already available infrastructure~\cite{winston2010alternate}.
In this regard, technologists around the globe
have pointed to traffic lights optimizations 
as a key ingredient for the development of more autonomous and intelligent designs to improve traffic dynamics 
\cite{brockfeld2001optimizing,palatella2013nonlinear,knospe2002single,kareem2011intelligent, tubaishat2007adaptive, omachi2009traffic, chavan2009design}.

The free flow of traffic may be interrupted by a number of events or devices, resulting in complex phenomena depending on particular situations  and their feedback on the drivers' behavior. Crossroads are such a case, for example, where the right-of-way may be defined by the priority to the right rule~\cite{abramson2013}, signaled by yield or stop signs, or by traffic lights. Traffic lights (TL) operation is determined by
the characteristics times linked to the cycle red-yellow-green. 
The effects of this periodicity on the traffic flow is 
a non trivial phenomenon, since when the system is 
close to the jam state it  becomes highly unstable 
\cite{kerner1997asymptotic, kerner1998experimental, kerner2012physics,nagatani2000density},
and small perturbations may lead to a full congestion.
In this context, the idea of this work is to characterize 
the effect of TL performance  on traffic flow, taking also into account behavioral aspects of the drivers. 
We propose a model base on a cellular automaton 
\cite{nagel1992cellular,nagel1994life,nagel1995emergent,cheybani2000nondeterministic,jia2011traffic}
where we have incorporated a system of traffic lights. Also, the behavior of drivers will be characterized by two parameters that account for their reactions when driving on the road. 

\section{The model}

The model we propose here is based on the work 
of Nagel and Paczuski \cite{nagel1995emergent}, where the 
representation of the traffic along a one direction
road is a one-dimensional array of length $L$. Vehicles move at discrete time intervals, with integer velocities in the range $[0, v_{max}]$. Each vehicle occupies only one site
and each site can host only one car or be empty. A car is in a free flow state when
it is travelling at $v_{max}$, otherwise it is in a jam state. In order to change velocities, i.e.~the number
of cells to move forward, cars evaluate their current velocity and the distance to  
the vehicle ahead,  $\Delta$. Any change must obey  the following rules:
\begin{itemize}
\item[a] If a car is in free flow and  $\Delta \geq v_{max}$ then the car keeps is current velocity.
\item[b] If the previous condition is not satisfied,  the car either keeps or changes its current velocity according to the following rules:
\begin{itemize}
    \item[b.1] Acceleration. If $\Delta \geq v+1$, then with probability $1-p$ it
accelerates: $v \rightarrow v+1$. Otherwise it keeps the same velocity.
    \item[b.2] Deceleration. If $\Delta \leq v-1$, then the car decelerates in a 
    prevention-based attitude: with probability $q$, decelerates to $v \rightarrow \max(\Delta-1, 0)$,
    and with complementary probability $1-q$, decelerates 
    to $v \rightarrow \Delta$.
\end{itemize}
\end{itemize}
Once the new velocities have been chosen, each car moves $v$ sites forward.

Note that in Ref.~\cite{nagel1995emergent} $p=q=1/2$ are fixed
parameters of the model. In this work we want to link these two parameters to behavioral responses of the drivers, so we will let them adopt different values within the interval $[0,1]$. 
The parameter $q$ is associated to the probability of adopting a prevention-based
attitude (``over-brake''), whereas $p$ can be interpreted  as the probability that a driver keeps an extra safe distance to the car ahead.

We choose a system with  open boundary conditions. Vehicles enter the road from
the left and get out from the right. Every car enters with velocity $v=0$,
and their entry rate is defined by a probability $\alpha$, such that 
if $\alpha =0$ no cars enter the road, whereas if $\alpha =1$ each time the leftmost cell is empty a car enters the system. Cars leaving the road, on the other hand, leave as if they had a free road ahead.

The traffic lights are modeled considering only two states: green and red, indicating \emph{stop} and \emph{go} respectively, as usual. The TL is placed at distance $X$ from the road entry, and its periodic dynamics is characterized by $\tau = t_R + t_G$, where $t_R$ and $t_G$ are the duration of the red and green states respectively, such that:
\begin{equation}
\begin{split}
    t_R &= \gamma\, \tau, \\
    t_G &= (1- \gamma) \tau,
\end{split}
\end{equation}
where $\gamma \in [0,1]$.

\begin{figure*}[!t]
\centering
\includegraphics[width=1\textwidth]{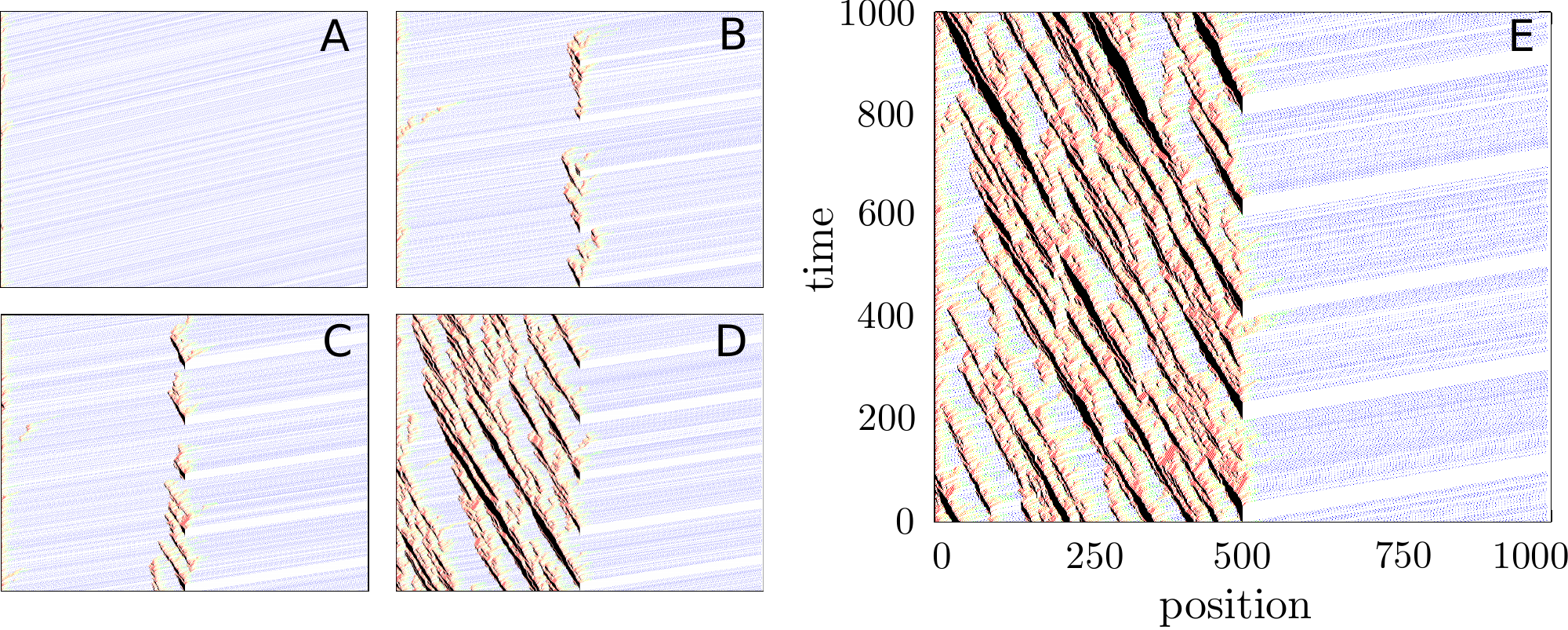}
\caption{
Representation of the traffic state in the plane $(x,t)$, for different values of $\gamma$.
Panel A: without TL, free flow. Panels B and C: $\gamma=0.1$ and $\gamma=0.12$ respectively,
the red light time does not produce important perturbations
in the system, just short lines of cars near the TL position.
Panels D and E: $\gamma=0.14$ and $\gamma=0.16$ respectively, the system clearly exhibits jam states before the TL position. The color scale in the plot indicate the cars velocities,
blue:  $v=5$ (maximum velocity), green: $v=4$, yellow: $v=3$, orange: $v=2$; red: $v=1$,
black: $v=0$ (car stopped), white: empty site.
}
\label{matriz}
\end{figure*}

Figure~\ref{matriz} shows the traffic flow in the plane $(x,t)$. We plot the positions of the cars 
as a function of time, with colors representing the corresponding velocity. In this case, the parameters of the system were set at
$p=q=1/2$, $\tau = 200$, $\alpha = 1$, $v_{max}=5$, $L= 1000$ and $X = 500$. 

Panel A in the figure shows the evolution of the system for $\gamma=0$. We can see the cars entering the road with $v=0$ (black), accelerating until reaching $v_{max}$ (blue) and crossing the road without obstacles. When the traffic is in free flow, the position of the cars traces a diagonal in the plane $(x,t)$, with slope $1/v_{max}$. This is what characterizes the \emph{free flow} state. Panels B and C, show the behaviour of the system for small values of $\gamma$ (short red state of the TL). We observe cars stopping at $X$ and the formation of short waiting lines (partial congestion) induced by the red light, and free flow ahead of the TL. 
In panel D we can see a stronger effect of the traffic light on the system. At $\alpha = 0.14$, i.e.~14\% of the cycle in the red state, the system goes to a \emph{full jam} state. In this situation it is possible to observe Nagel waves \cite{nagel1992cellular}, which is a phenomenon characterized by  the congestion moving backwards on the road and forward in time. These are shock waves that propagate the jam state to the entire system.
In this state, long lines are formed at the TL and consequently the cars near the end must wait more than one cycle to cross the TL. This simple example puts in evidence how the perturbation linked to the lights cycle may induce  the transition from the free flow to a full jam state.
In the following section, we analyze the problem from the perspective of a phase transition.
We recall that the behavioral aspects have  not been considered so far.

\begin{figure*}[!t]
\centering
\includegraphics[width=1\textwidth]{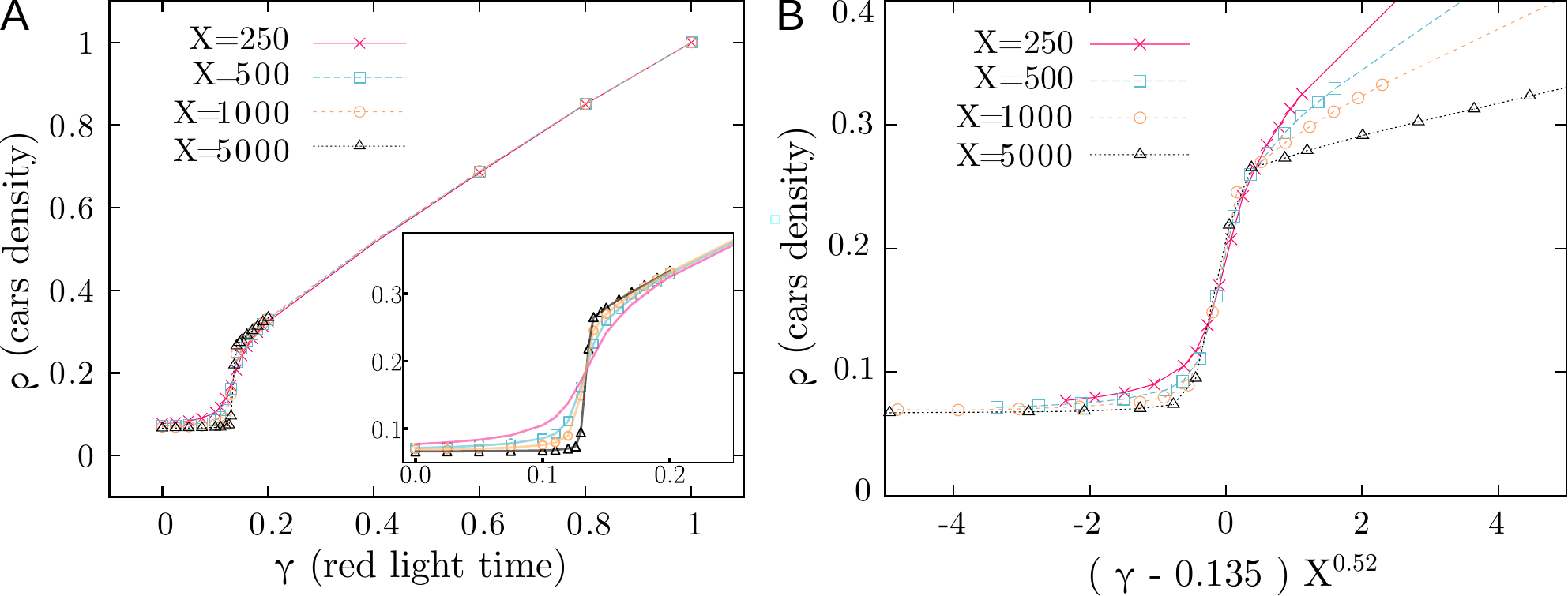}
\caption{
Density as a function of the red light parameter.
In panel A, the main plot shows the curves for different values of $X$,
and the inset the same in the vicinity of the transition.
Panel B shows curves collapse for the different values of $X$ when
we express the density as function of the critical 
parameters $\rho= \Phi((\gamma - \gamma_c) X^{\xi_c})$.
For both plots the other parameters are set to $\tau=200$, $p=q=1/2$ and $\alpha=1$.
The total length of the road was chosen as $L= X + 200$, leaving an additional distance behind the TL
for the cars to evolve in free flow until they leave the system.
}
\label{densidad}
\end{figure*}

\section{Results} 

In this section we present the results
obtained from extensive numerical simulations
of the model proposed above.
We study the state of the traffic 
as a function of the red light parameter $\gamma$.
To do so we measure the vehicle density, $\rho$, from the entry to the TL position, $X$,
as a function of $\gamma$.
We proceed as follows. First, we perform an initialization allowing the system to evolve 
during 10 cycles of $\tau$, since by inspection we observed that this is enough to eliminate  
any correlation linked to the initial conditions.
Then, at each time step during ten additional cycles, we measure the density as the number of cars
divided the number cells, obtaining an average density $\rho$ as a function of 
the parameter $\gamma$.
Figure \ref{densidad} panel A shows the relation 
$\rho$ vs. $\gamma$ for different values 
of $X$. 
We can see that the value of $\rho$ remains almost  constant  for small values of $\gamma$, indicating a
state of free flow; whereas for values close to $\gamma \approx 0.13$, the density exhibits 
an abrupt change, which seems to indicate a first order transition to a jam state~\cite{echenique2005dynamics}. 
After this transition $\rho$ increases linearly with $\gamma$ 
until reaching the full jam state at $\gamma=1$.
Besides, the transition becomes sharper as $X$ increases, which suggests  the existence of a function 
$\Phi$ such that
\begin{equation} 
\rho = \Phi\left((\gamma - \gamma_c)X^{\xi_c}\right),
\end{equation}
describing an universal behavior of the density close to the transition region.
By performing a nonlinear fit of the curves we calculated the critical parameters:
$\gamma_c= 0.135 \pm 0.003$ and $\xi_c = 0.52 \pm 0.05$.
We used these values to plot the collapsed curves in Fig.~\ref{densidad} panel B.

\begin{figure*}[!t]
\centering
\includegraphics[width=1\textwidth]{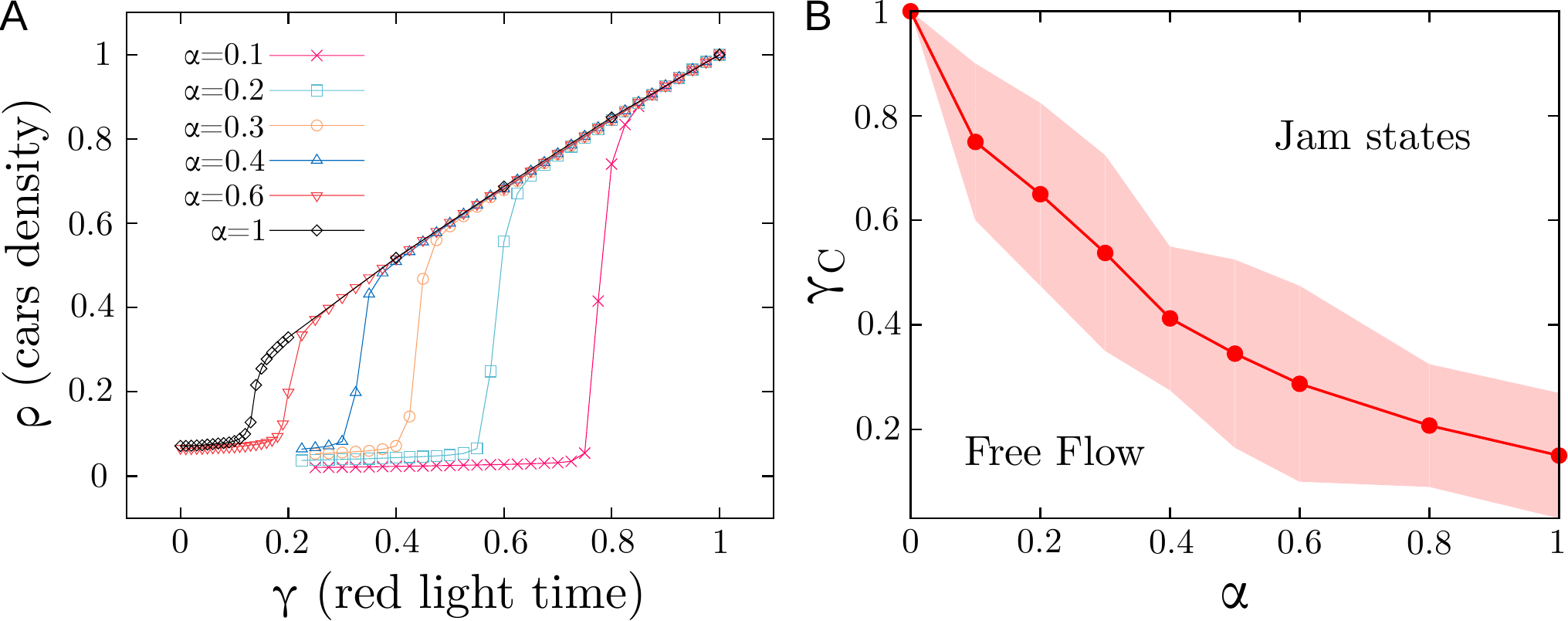}
\caption{
Panel A: relation $\rho$ vs. $\gamma$ for different values of 
the entry probability $\alpha$.
The rest of the parameters are fixed at $\tau=200$, $p=q=1/2$ and $X=500$.
Panel B: phase diagram of free flow and jam states,
in the space of the parameters $\alpha$ and $\gamma_c$.
The shaded region indicates where the system
becomes unstable and the density starts to increase.
}
\label{alpha}
\end{figure*}


We turn now our attention to the effects linked to the entry flow.
This is controlled by the parameter  $\alpha$, which rules
the  probability that, at each time step and under suitable conditions, a car enters the road from the left. Figure~\ref{alpha} shows the result 
for the case $X=500$.
We observe that the value of  $\gamma_c$ has a strong dependence on $\alpha$, revealing a relationship  between the number of cars in the road and the critical red light time, responsible of driving  the system to a full jam state. To shed light upon this, 
we estimate $\gamma_c$ for every curve in Fig.~\ref{alpha} panel A, and plot the relation $\gamma_c$~vs.~$\alpha$ in Fig.~\ref{alpha} panel B. The band shaded red in the plot indicates the parameters'  region  where the system changes from free flow to the jam state. 

\begin{figure*}[!t]
\centering
\includegraphics[width=0.6\textwidth]{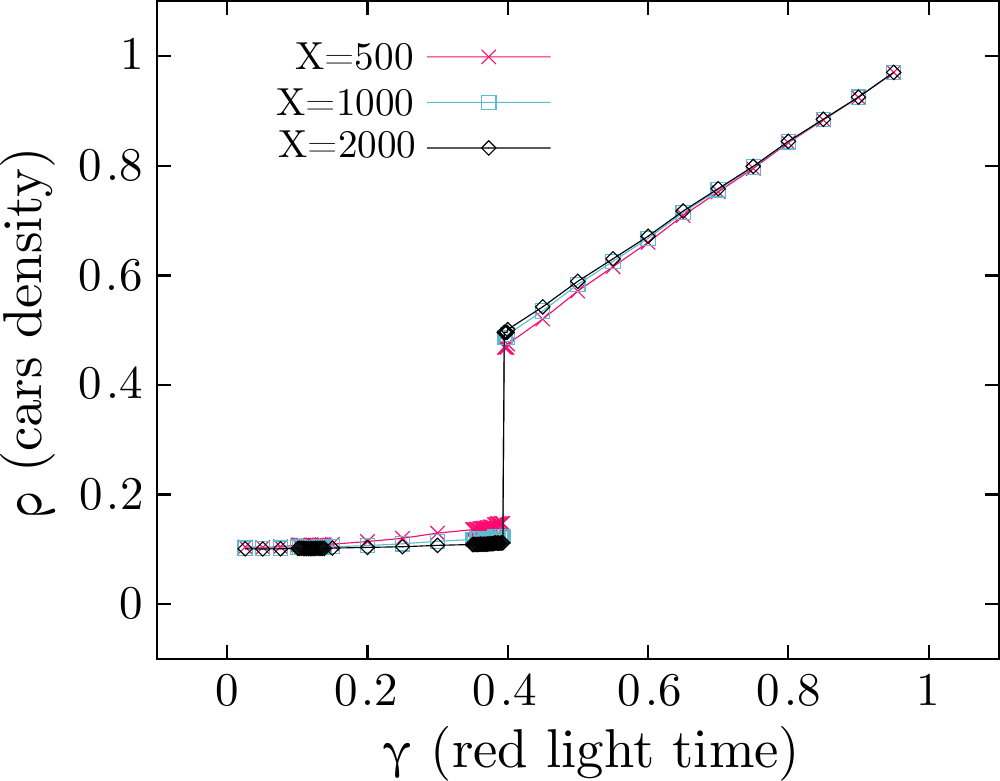}
\caption{
The deterministic system. 
The parameters are set at $\tau=200$, $\alpha=1$ and $p=q= 0$.
We can see an abrupt change in the density at
$\gamma=0.4$, independently of system size.
}
\label{determinista}
\end{figure*}

Finally, we study the role of the parameters 
$p$ and $q$ which, as we said before, rule
the drivers' behavior at a microscopic level but lead to macroscopic  
effects on traffic. 
In the simulations presented below we adopted  fixed values of the parameters $\tau=200$, $\alpha=1$ and $X=500$. We analyze three different scenarios. First we study the deterministic case, 
i.e.~$p = q = 0$, when the drivers behave always in the same manner, and therefore the noise associated with human behavior vanishes.
In this case, the transition to the jam states happens near $\gamma \sim 0.4$
(see Fig.~\ref{determinista}). Since there is no stochasticity, the origin of this transition is just an overload of cars on the road. At the critical value of $\gamma$, the green state duration is not long enough to clear the line formed during the red one and consequently the system becomes jammed.

Then we study the case with fixed $q$ and different values of $p$, with the idea of revealing the effects of keeping a safe distance between cars. In Fig.~\ref{reglas} panel A we can see that, as $p$ decreases, the transition moves towards larger values of $\gamma$, indicating that  the transition is delayed. Similarly, if we fix the parameter $p$ and vary $q$ (in order to study the effect of keeping a  safe velocity), we can see in Fig.~\ref{reglas} panel B the same effect. However, in this last case the transition appears to be less abrupt. Note in particular that, for $q=0.75$, the transition becomes so blurred that we are not able to clearly distinguish either the phase of free flow or the value $\gamma_c$. We can conclude that  the human behaviour produces local  effects that spread non trivially through the entire system.

In the following section, we discuss these results in more detail.


\section{Discussion}

We present here a more thorough analysis of the main findings. Let us first address the results presented in   Fig.~\ref{densidad}, where we 
studied the relation between the red light duration and vehicular density. We can observe that for $\gamma \approx 0$, the density reaches an approximate constant value $\rho \approx 0.07$ cars per site.
This value agrees with the one reported by Nagel in Ref.~\cite{nagel1992cellular},
for systems with open boundary conditions.

In the absence of TL  the system  would remain  in a free flow state; however the red 
light makes the vehicles stop during a time $t_R$, which causes the density on the road before the TL to increase and induce instabilities that lead the traffic to a phase change.
The critical density $\rho \approx 0.1$ cars per site reported in Ref.~\cite{nagel1992cellular}
for the case with periodic boundary conditions, agrees with the one observed in our simulations. 
Once the density on the road reaches that threshold, corresponding to $\gamma \approx 0.13$, any increase in $\gamma$ will lead the system to the jam state. Therefore, we can conclude that the effect produced by the TL agrees with other studies reported in the literature. 

Another interesting result is the relation between $\rho$ and $\gamma$ for varying 
 entry flows $\alpha$ (see Fig.~\ref{alpha}). We can observe that there is a shift in $\gamma_c$ towards 
larger values, as $\alpha$ decreases. Clearly, for a system with an entry flow $\alpha =0.1$,
for instance, the red light duration needed for the road to reach the critical density will be longer than the one needed for $\alpha= 1$. Interestingly, the behavior of the density after the transition
seems to be independent of $\alpha$. In this conditions the backwards propagating wave of jammed traffic reaches the entry, minimizing any macroscopic effect of the inflow of new cars at the entry. 


\begin{figure*}[!t]
\centering
\includegraphics[width=1\textwidth]{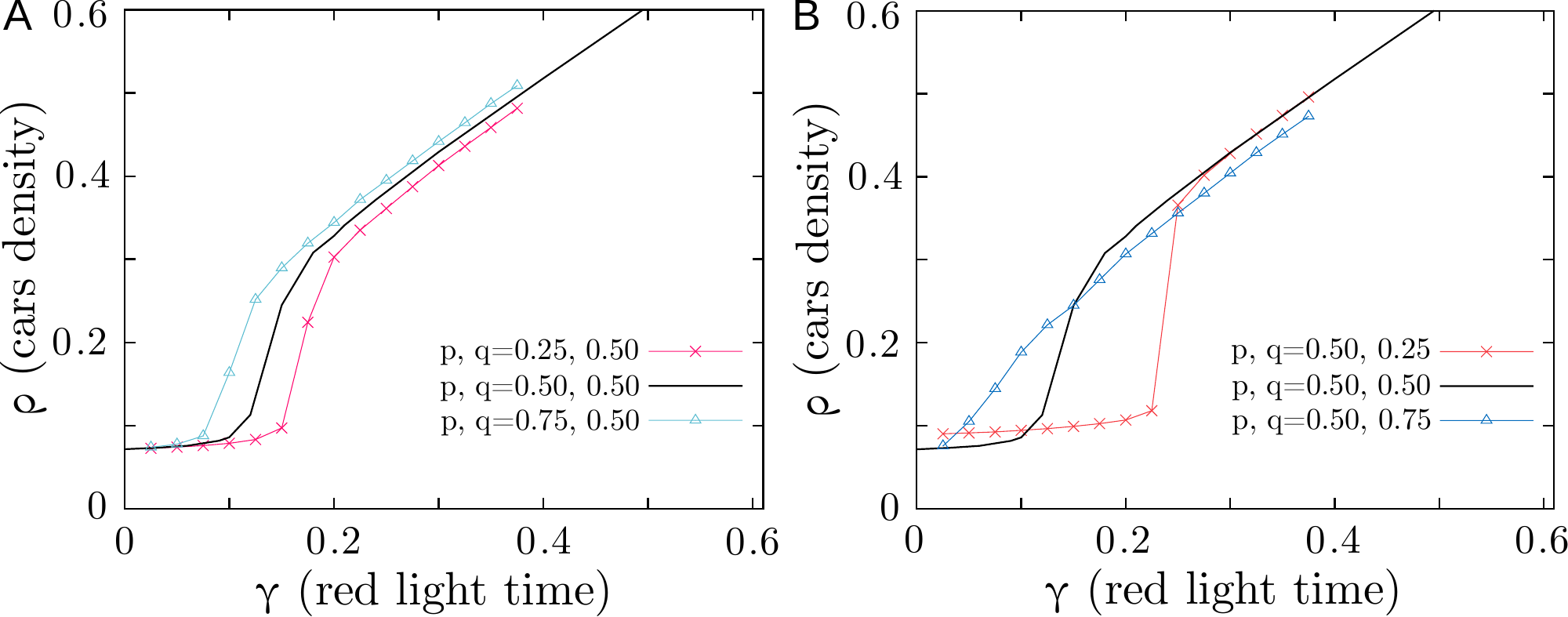}
\caption{
The effect of the drivers' behavior (associated to  $p$ and $q$)
on the value of $\gamma_c$.
Panel A: fixed $q$ and varying $p$.
Panel B: fixed $p$ and varying $q$.
The other parameters are $\tau=200$, $\alpha=1$ and $X=500$.
}
\label{reglas}
\end{figure*}

Finally, we have the results associated with the behavioral aspects of the drivers in the model, controlled by the parameters $p$ and $q$. We studied
three cases: (i) The deterministic case, where the 
drivers act always in the same way; (ii) the case for
different values of $p$, which is the probability of 
keeping a safe distance; 
and (iii) the case for different values of $q$, which
is the probability of keeping a safe velocity. 

In the deterministic case (i), the TL cycle 
becomes critical at $\gamma= 0.4$, as shown in Fig.~\ref{determinista}. As we said before, this transition is not caused by fluctuations but by an overload of the road capacity.
Namely, the green light time corresponding to $\gamma = 0.4$ is not long enough to clear the waiting  line and  thus the traffic collapses. Since in this case we do not have the noise associated to the drivers behavior, we can estimate the value of $\gamma_c$ as follows.
Let us define $J_{in}$ as the vehicle entry rate per unit of time, under the constraint that a new car enters the road if and only if the first site of the road is empty. Then:
\begin{equation}
J_{in} = (1 - J_{in})\alpha~~\Rightarrow~~J_{in}=\frac{\alpha }{1+\alpha },
\end{equation}
since $1-J_{in}$ is the fraction of time when the entry site is empty and $\alpha$ is the entry probability. Moreover, the exit flow through the TL must satisfy:
\begin{equation}
\avg{J_{out}} =  (1-\gamma) \frac{v_{max}}{1+v_{max}}, 
\end{equation}
corresponding to the flow of cars exit at an an infinite jam (see Ref. 
\cite{nagel1995emergent} section II). Then the transition is produced when the entry
flow balances the exit flow, $\avg{J_{IN}}=\avg{J_{OUT}}$, resulting in:
\begin{equation}
\gamma_c = 1 - 
\left ( \frac{\alpha}{1+\alpha}     \right )
\left ( \frac{1 + v_{max}}{v_{max}} \right ).
\label{eq:gamma_C}
\end{equation}
In particular, by replacing in Eq.~(\ref{eq:gamma_C}) 
the values $\alpha =1$ and $v_{max}=5$, we obtain
$\gamma_c=0.4$, which is consistent with the value observed  in the simulations for the deterministic case shown in Fig.~\ref{determinista}.

In case (ii) (see Fig.~\ref{reglas} A),  
when we reduce the value of $p$, the cars  will tend  to get closer to the 
one ahead more frequently. This has the effect of a reduction of the noise
in the system and consequently $\gamma_c$ approaches 
the critical value of the deterministic case.
On the other hand, if we increase $p$, the cars  will tend to leave an empty site 
between them and the one ahead. Therefore, contrary to the previous case, this extremely cautious attitude cause smaller velocities in the system, which attempts against the fast exit of the queue,
which leads to a reduction of the value of $\gamma_c$.
This phenomenon makes the system to create \textit{virtual
traffic lights}, where the safe attitude leads to stops governed by caution rather than TL. In turn, this  makes the density of cars to increase for smaller values of $\gamma$ with respect 
to the original system. 

Finally in case (iii), when $q$ is small the cars tend to accelerate as much as possible. 
The curve for $q=0.25$ in Fig.~\ref{reglas} B shows a shift in $\gamma_c$ towards larger values.  
As in the previous cases this is related to the reduction of noise. 
On the other hand, for larger values of $q$, the cautious attitude cause lower velocities in the system, and a further increase of the car density independently of the red light times.

Summarizing, when we change the parameters $p$ and $q$, there is a compromise between the noise level 
and the slowing down caused by the drivers safe attitude.
Even though the extremes values of these parameters are not observed in real traffic systems, 
this analysis becomes useful to understand the underlying processes. 
Moreover, as autonomous vehicles become more common, drivers that do not abide by the cautious behavior of most human drivers will play a significant role in the dynamics of traffic.

Notice that the possibility of controlling
$\gamma_c$ becomes extremely relevant when we have a
traffic systems with more than one road.
For instance at a crossroads ruled by traffic lights, the red light time of one road corresponds to the green one of the other, and vice versa. Therefore, signaled crossroads can be
thought as coupled systems where the optimization
of $\gamma_c$ is an important tool to guarantee the common well-being in both roads.

\section{Conclusion}

In this work we studied the effect of a periodic 
traffic light system in a one-way road by
performing numerical simulations based on the model 
of cellular automata proposed in Ref.~\cite{nagel1995emergent}, with open boundary conditions.
In this frame we observed that: (i)  without traffic lights, 
the system reaches an stationary state
of free flow, with low and constant mean density; and
(ii) with traffic lights, the effect of the red light time
produces a density increase and consequently 
instabilities and fluctuations on the traffic.
We proposed an analysis based on the effect caused
by the variation of the main parameters of the system.

Firstly, we observed that, when the red light time is long
enough, instabilities lead the system to a phase
transition from free flow to jam state.
In this conditions, the density curves as a function of
the red light time exhibit an abrupt change which
characteristic of a first order phase transition. Moreover, by varying the road
length before the TL, we observed a finite-size effect, characterized by a function 
$\rho = \Phi\left(\gamma - \gamma_c)X^{\xi_c}\right)$ which 
describes a universal behavior of the vehicle density around the transition.

Secondly,
we studied the effect of varying the entry rate to the
road, and observed a shift in $\gamma_c$, which
we summarized in the phase diagram of Fig.~\ref{alpha} panel B, which displays
two main regions: free flow and jam states.
This is an interesting result from a technical 
point of view, since it might be used as input data in 
control systems for intelligent traffic lights. 

Thirdly, we focused on the effect of drivers'
behavior, as characterized by different
combinations of the parameters $p$ and $q$, ruling the drivers' 
trend to behave safely. In the deterministic case 
we observed a transition to jam states not caused by
fluctuation but by a road overload capacity instead,
where the green light time is not long enough to evacuate 
the line at the TL formed during the red light phase. 
For this case, 
by performing an analysis of the flow balance, 
we estimated analytically the value of $\gamma_c$,
complementing the one observed in our numerical simulations.
On the other hand, we analyzed individually 
the effect of $p$ and $q$. We observed in both cases that extremely 
cautious attitudes lead to delays in the traffic systems,
where the drivers' behavior create 
locally an effect similar to a TL, which globally produces 
a decrease in the value of $\gamma_c$,
worsening the state of the traffic.

In summary,  in this work we have analyzed a simple traffic situation, consisting in a single lane road and a traffic light.
By adjusting different parameters accounting for the relative periods of red and green lights, as well as different behavioral profiles of the 
drivers, we found interesting macroscopic effects on the vehicle flow.
We have unveiled the negative effects that small perturbations, originated by the presence of the traffic lights have on this flow and explored the 
optimal range of parameters to minimize them. 
These results can be extended to the study of more complex traffic systems, with
different roads topology, by adding data from flow measurements, or proposing other behavior patterns
more similar to the ones observed in real traffic situations.
The rationale behind these actions is to optimize traffic lights performances and to propose improvements on the system.


\section*{Acknowledgements} 
We acknowledge enlightening discussions with Alexandre Nicolas.
Partial support through the following grants: CONICET PIP 112-2017-0100008 CO, UNCUYO SIIP 06/C546,
CONICET (PIP 112 20150 10028), FonCyT (PICT-2017-0973), SeCyT–UNC (Argentina) and MinCyT Córdoba
(PID PGC 2018).


\end{document}